\documentclass[aps,twocolumn,prl,showpacs,showkeys,preprintnumbers,superscriptaddress,nobibnotes,floatfix,longbibliography,notitlepage,nofootinbib]{revtex4-2}

\usepackage{amsmath}
\usepackage{amsfonts}
\usepackage{amssymb}
\usepackage{graphicx}
\usepackage[colorlinks=true,allcolors=blue]{hyperref}
\usepackage{relsize}
\usepackage{gensymb}
\usepackage{fontawesome}
\usepackage[capitalize]{cleveref}
\usepackage[dvipsnames,table]{xcolor}
\usepackage{lettrine}
\usepackage{fancyhdr} 
\usepackage{poemscol} 
\usepackage{comment}

\input Zallman.fd

\usepackage{siunitx}

\DeclareSIUnit \s {\second}
\DeclareSIUnit \ns {\nano\second}
\DeclareSIUnit \mus {\micro\second}
\DeclareSIUnit \ms {\milli\second}
\DeclareSIUnit \MB {\mega\byte}
\DeclareSIUnit \GB {\giga\byte}
\DeclareSIUnit \TB {\tera\byte}
\DeclareSIUnit \PB {\peta\byte}
\DeclareSIUnit \Mbps {\mega\bit/\s}
\DeclareSIUnit \Gbps {\giga\bit/\s}
\DeclareSIUnit \Tbps {\tera\bit/\s}
\DeclareSIUnit \Pbps {\peta\bit/\s}
\DeclareSIUnit \kton {\kilo\tonne} 
\DeclareSIUnit \kt {\kilo\tonne}
\DeclareSIUnit \kty {\kilo\tonne-\year}
\DeclareSIUnit \Mt {\mega\tonne}
\DeclareSIUnit \eV {\electronvolt}
\DeclareSIUnit \keV {\kilo\electronvolt}
\DeclareSIUnit \MeV {\mega\electronvolt}
\DeclareSIUnit \GeV {\giga\electronvolt}
\DeclareSIUnit \TeV {\tera\electronvolt}
\DeclareSIUnit \PeV {\peta\electronvolt}
\DeclareSIUnit \EeV {\exa\electronvolt}
\DeclareSIUnit \sr {sr}
\DeclareSIUnit \m {\meter}
\DeclareSIUnit \cm {\centi\meter}
\DeclareSIUnit \nm {\nano\meter}
\DeclareSIUnit \in {\inchcommand}
\DeclareSIUnit \km {\kilo\meter}
\DeclareSIUnit \kV {\kilo\volt}
\DeclareSIUnit \kW {\kilo\watt}
\DeclareSIUnit \MW {\mega\watt}
\DeclareSIUnit \MHz {\mega\hertz}
\DeclareSIUnit \mrad {\milli\radian}
\DeclareSIUnit \year {years}
\DeclareSIUnit \POT {POT}
\DeclareSIUnit \sig {$\sigma$}
\DeclareSIUnit\parsec{pc}
\DeclareSIUnit\lightyear{ly}
\DeclareSIUnit\foot{ft}
\DeclareSIUnit\ft{ft}
\DeclareSIUnit \ppb{ppb}
\DeclareSIUnit \ppt{ppt}
\DeclareSIUnit \samples{S}
\DeclareSIUnit \pe{PE}
\DeclareSIUnit \GeVmwe{GeV/mwe}
\DeclareSIUnit \mwe{mwe}

\newcommand{\enu}{\E_\enu}

\begin{document}

\title{Tau Appearance from High-Energy Neutrino Interactions}

\author{A.~Garcia~Soto}
\email{alfonsogarciasoto@fas.harvard.edu}
\affiliation{Department of Physics \& Laboratory for Particle Physics and Cosmology, Harvard University, Cambridge, MA 02138, USA}
\affiliation{Instituto de Física Corpuscular (IFIC), Universitat de València (UV), 46980 Paterna, València, Spain.}

\author{P.~Zhelnin}
\email{pzhelnin@g.harvard.edu}
\affiliation{Department of Physics \& Laboratory for Particle Physics and Cosmology, Harvard University, Cambridge, MA 02138, USA}

\author{I.~Safa}
\email{isafa@fas.harvard.edu}
\affiliation{Department of Physics \& Laboratory for Particle Physics and Cosmology, Harvard University, Cambridge, MA 02138, USA}
\affiliation{Department of Physics \& Wisconsin IceCube Particle Astrophysics Center, University of Wisconsin-Madison, Madison, WI 53706, USA}

\author{C.~A.~Arg{\"u}elles}
\email{carguelles@fas.harvard.edu}
\affiliation{Department of Physics \& Laboratory for Particle Physics and Cosmology, Harvard University, Cambridge, MA 02138, USA}

\date{\today}

\begin{abstract}
High-energy muon and electron neutrinos yield a non-negligible flux of tau neutrinos as they propagate through Earth.
In this letter, we address the impact of this additional component in the PeV and EeV energy regimes for the first time.
Above $\SI{300}\TeV$, this contribution is predicted to be significantly larger than the atmospheric background, and it alters current and future neutrino telescopes' capabilities to discover a cosmic tau-neutrino flux.
Further, we demonstrate that Earth-skimming neutrino experiments, designed to observe tau neutrinos, will be sensitive to cosmogenic neutrinos even in extreme scenarios without a primary tau-neutrino component. 
\end{abstract}

\maketitle

\begin{poem}
\begin{stanza}
\textit{To tau or not to tau, that is the question. \verseline
Whether 'tis nobler underground to measure, \verseline
tracks and cascades of outrageous flavors,\verseline
or to hunt taus against a sea of troubles \verseline
and by good fortune get them.}
\end{stanza}
\end{poem}
\vspace{-0.50cm}
\lettrine[nindent=0em]{E}{arth} is continuously bombarded by neutrinos of extraterrestrial origin~\cite{Vitagliano:2019yzm}.
The study of astrophysical neutrinos has proven fruitful over the years~\cite{Arnett:1987iz,Goldman:1987fg,Manohar:1987ec,Lattimer:1988mf,Raffelt:1987yt,Turner:1987by,Bethe:1990mw} --- with the discovery of solar neutrinos proving that nuclear fusion is the fuel of stars~\cite{Raffelt:1999tx,Bahcall:2004pz}.
Later, studies of solar neutrino flavor structure led to the discovery of neutrino oscillations~\cite{McDonald:2016ixn}, which to this day remains a major unresolved mystery in the Standard Model.
More recently, the IceCube Neutrino Observatory~\cite{Aartsen:2016nxy} at the South Pole has measured a spectrum of astrophysical neutrinos of extragalactic origin --- providing evidence of a predominantly muon- and electron-flavored neutrino flux in the TeV to PeV energy range~\cite{Aartsen:2013jdh,IceCube:2020wum,Stettner:2019tok}.
However, much remains unknown about the properties and flavor composition of this flux.
For most source emission scenarios, the expected flavor composition at Earth is approximately $\left(1:1:1\right)$ due to standard neutrino oscillations.
Extreme deviations from this expectation can happen under several Beyond the Standard Model (BSM) scenarios~\cite{Pakvasa:2012db, Pagliaroli:2015rca, Huang:2015flc, Shoemaker:2015qul, Bustamante:2016ciw, Denton:2018aml, Arguelles:2019rbn, Abdullahi:2020rge, Bustamante:2020niz,Rasmussen:2017ert,Arguelles:2019rbn}, making flavor measurements a powerful tool in new physics searches.
In particular, several BSM scenarios predict significant attenuation of cosmic tau neutrinos~\cite{Brdar:2016thq,Pakvasa:2012db, Pagliaroli:2015rca, Huang:2015flc, Shoemaker:2015qul, Bustamante:2016ciw, Denton:2018aml, Abdullahi:2020rge, Arguelles:2019rbn, Bustamante:2020niz,Rasmussen:2017ert,IceCube:2021tdn}.
It is therefore fundamental to constrain the relative contribution of this flavor.
Additionally, the detection of high-energy tau neutrinos is crucial in neutrino astronomy for two main reasons:
1) high-energy tau neutrinos are a smoking gun signature of astrophysical neutrinos as they are rarely produced in the atmosphere~\cite{Garzelli:2015psa,Gauld:2015kvh,Gauld:2015yia,Bhattacharya:2016jce,Garzelli:2016xmx,Goncalves:2006ch,Enberg:2008te,Bulmahn:2010pg} and 2) measuring the relative fraction of neutrino flavors in the cosmic flux provides information about the sources and how these neutral leptons propagate over astronomical distances~\cite{Rachen:1998fd, Bento:1999bb, Athar:2000yw, Barenboim:2003jm, Beacom:2003nh, Anchordoqui:2003vc, Kashti:2005qa, Xing:2006uk, Kachelriess:2006fi, Rodejohann:2006qq, Mena:2006eq, Lipari:2007su, Hwang:2007na, Pakvasa:2007dc, Esmaili:2009dz, Choubey:2009jq, Lai:2009ke, Hummer:2010ai, Mehta:2011qb, Fu:2012zr, Vissani:2013iga, Lai:2013isa, Mena:2014sja, Fu:2014isa, Bustamante:2019sdb,Arguelles:2015dca,Bustamante:2015waa,Song:2020nfh}.
However, at present the tau-neutrino component lacks a large sample size to make \textit{bona fide} conclusions~\cite{IceCube:2015gsk,IceCube:2018pgc}.

The interaction of tau neutrinos in Earth's surface produces a detectable extensive air shower (EAS) from the resulting tau-lepton decay in the atmosphere~\cite{Fargion:2000iz,Bertou:2001vm,Feng:2001ue}.
This signature will be used by next-generation neutrino experiments to detect ultra-high-energy neutrinos~\cite{Kotera:2021hbp,Venters:2019xwi,Wissel:2020sec,Wang:2021zkm}.
In particular, a flux of cosmogenic neutrinos is produced at these energies as a result of the interaction of ultra-high-energy cosmic-rays (UHECRs) with the cosmic microwave background~\cite{1966JETPL...4...78Z,PhysRevLett.16.748}.
Although its existence is guaranteed, the characteristics of this flux are unknown and are intrinsically related to the primary composition and low-energy cutoff of UHECRs and the sources' redshift distribution~\cite{Heinze:2019jou,vanVliet:2019cpl,Groth:2021bub,AlvesBatista:2018zui,Romero-Wolf:2017xqe,Das:2018ymz,Moller:2018isk,Ahlers:2012rz}.
This puzzle cannot be solved solely by looking at cosmic rays or photons since they lose valuable information on their journey to Earth.
Therefore, any measurement of the cosmogenic neutrino flux would shed light on the nature of UHECRs.

In this letter, we study how neutrino propagation through Earth can affect the interpretation of tau-neutrino measurements.
Neutrinos above $\SI{10}\TeV$ will interact in Earth via deep inelastic scattering (DIS), modifying the neutrino flux at the detector~\cite{Gandhi:1995tf}. 
Electron and muon neutrinos produce charged particles that rapidly lose energy and do not yield high-energy neutrinos.
However, tau-neutrino interactions generate secondary neutrinos that carry a significant fraction of the primary energy~\cite{Ritz:1987mh,Halzen:1998be,Iyer:1999wu,Dutta:2000jv,Becattini:2000fj,Beacom:2001xn,Dutta:2002zc,Jones:2003zy,Yoshida:2003js,Bugaev:2003sw,Bigas:2008sw,Alvarez-Muniz:2018owm,Safa:2019ege}.
This process makes Earth effectively transparent to tau neutrinos and opaque to other flavors at high energies. 
However, we show in this letter that this picture is incomplete as high-energy electron or muon neutrinos can also produce tau neutrinos.
We prove that this effective-tau-appearance phenomenon has a significant and direct impact on the inferred astrophysical neutrino spectra and flavor measurements.

\begin{figure}[!h]
\centering
\includegraphics[width=0.49\textwidth]{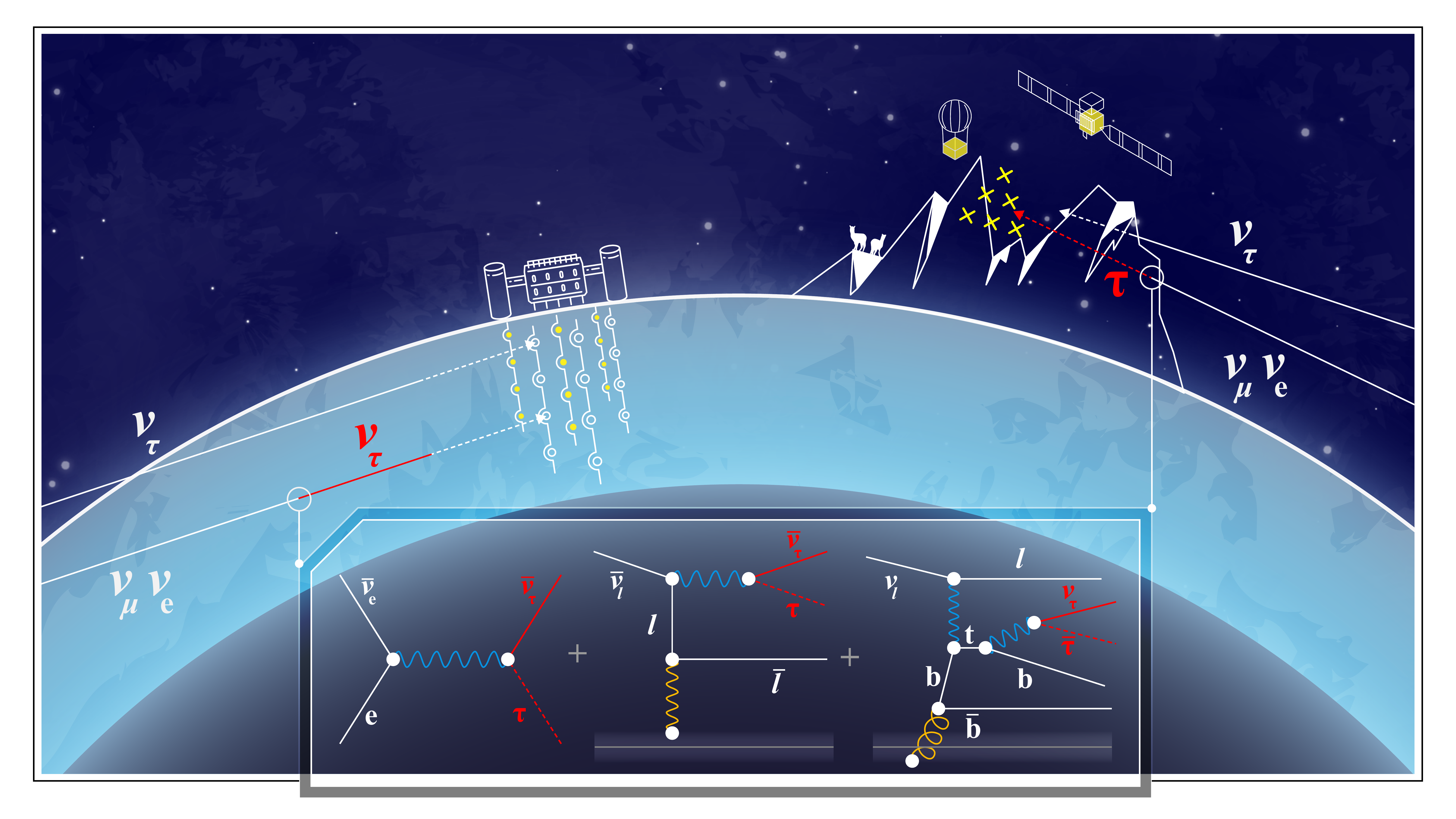}
\caption{\textbf{\textit{Artistic rendition of relevant interactions for tau appearance.}}
The Feynman diagrams summarize the muon- and electron-neutrino interactions that would produce a pair $\tau+\nu_\tau$: Glashow resonance (left), $W$-boson production (middle) and top-quark production (right). 
Ice- and water-Cherenkov detectors are represented on the center left. Earth-skimming and space-based observatories are shown on the center right.
}
\label{fig:scheme}
\end{figure}

\textbf{\textit{High-energy tau neutrino appearance ---}} For the first time, we have identified channels that would yield a significant contribution of secondary tau neutrinos as shown in~\cref{fig:scheme}.
The secondary contribution arises from channels that produce on-shell $W$ bosons, which decay $10\%$ of the time to $\nu_{\tau}+\tau$~\cite{ParticleDataGroup:2020ssz}.
The dominant interactions are neutrino-nucleus~\cite{Seckel:1997kk,Alikhanov:2015kla,Barger:2016deu,Zhou:2019vxt} or neutrino-electron~\cite{Glashow:1960zz,Gauld:2019pgt,Huang:2019hgs} $W$-boson production, which become relevant above PeV energies.
In the neutrino-nucleus interaction, when the energy transferred from a neutrino to a nucleon is above the top-quark mass, a $W$ boson can also be produced from the decay of this heavy quark to a $b+W$ pair~\cite{Barger:2016deu}.
The latter process is intrinsically related to the parton distribution functions (PDF) of the sea bottom in the nucleons.
Depending on the PDFs and mass scheme models, top-quark production accounts for $5-15\%$ of the total DIS cross sections above $\SI{100}\TeV$~\cite{Bertone:2018dse,Cooper-Sarkar:2011jtt,Cooper-Sarkar:2007zsa,Gluck:2010rw,Connolly:2011vc,Albacete:2015zra,Arguelles:2015wba,Goncalves:2010ay}.

The propagation of neutrinos through Earth is simulated using the~\texttt{NupropEarth} framework~\cite{Garcia:2020jwr}.
\texttt{NupropEarth} has the structure of a general purpose Monte Carlo event generator and therefore tracks the interactions of neutrinos as they travel through Earth on an event-by-event basis.
We have calculated the relevant neutrino differential cross sections and implemented them in  \texttt{GENIE}~\cite{GENIE:2021npt}.
In our Monte Carlo, tau energy losses are simulated with \texttt{PROPOSAL}~\cite{Koehne:2013gpa} using the ALLM97 parametrization to describe the photo-nuclear interactions~\cite{Abramowicz:1997ms,Butkevich:2001aw}.
The kinematics of the outgoing products in fully polarized tau decays~\cite{Hagiwara:2003di} is handled by \texttt{TAUOLA}~\cite{Davidson:2010rw}.
Finally, Earth's density profile follows the PREM model~\cite{Dziewonski:1981xy}.

\textbf{\textit{Diffuse cosmic neutrino fluxes ---}} 
The astrophysical neutrino flux has been characterized by the IceCube Collaboration using multiple complementary channels~\cite{Aartsen:2013jdh,Aartsen:2014gkd,Aartsen:2014muf,Aartsen:2015zva,Aartsen:2017mau,IceCube:2018pgc,IceCube:2020acn,IceCube:2020wum,Abbasi:2021viw}. 
As a benchmark scenario, motivated by the Fermi acceleration mechanism~\cite{gaisser2016cosmic}, these analyses model the astrophysical neutrino flux as an unbroken power law in energy.
The results of these analyses agree within errors, and the best-fit spectral index ranges from $2.37 \to 2.89$.
The softest spectral index is measured using high-energy starting events (HESE) with deposited energy above $\SI{60}\TeV$~\cite{IceCube:2020wum}, while the hardest spectrum is measured when selecting for throughgoing muons in the northern sky~\cite{Stettner:2019tok,Abbasi:2021viw}.
Both analyses set constraints on the fraction of muon neutrinos, whereas similar signatures of electrons and taus in the detector give rise to a degeneracy between the other two flavors.
The smoking-gun signature of tau neutrinos is two separated energy depositions, known as a {\it double bang}. 
These have recently been detected~\cite{stachurska_juliana_2018_1301122}; still, the sample size does not yield significant evidence of a tau-neutrino component~\cite{IceCube:2020abv}.

\begin{figure}[!h]
\centering
\includegraphics[width=0.4\textwidth]{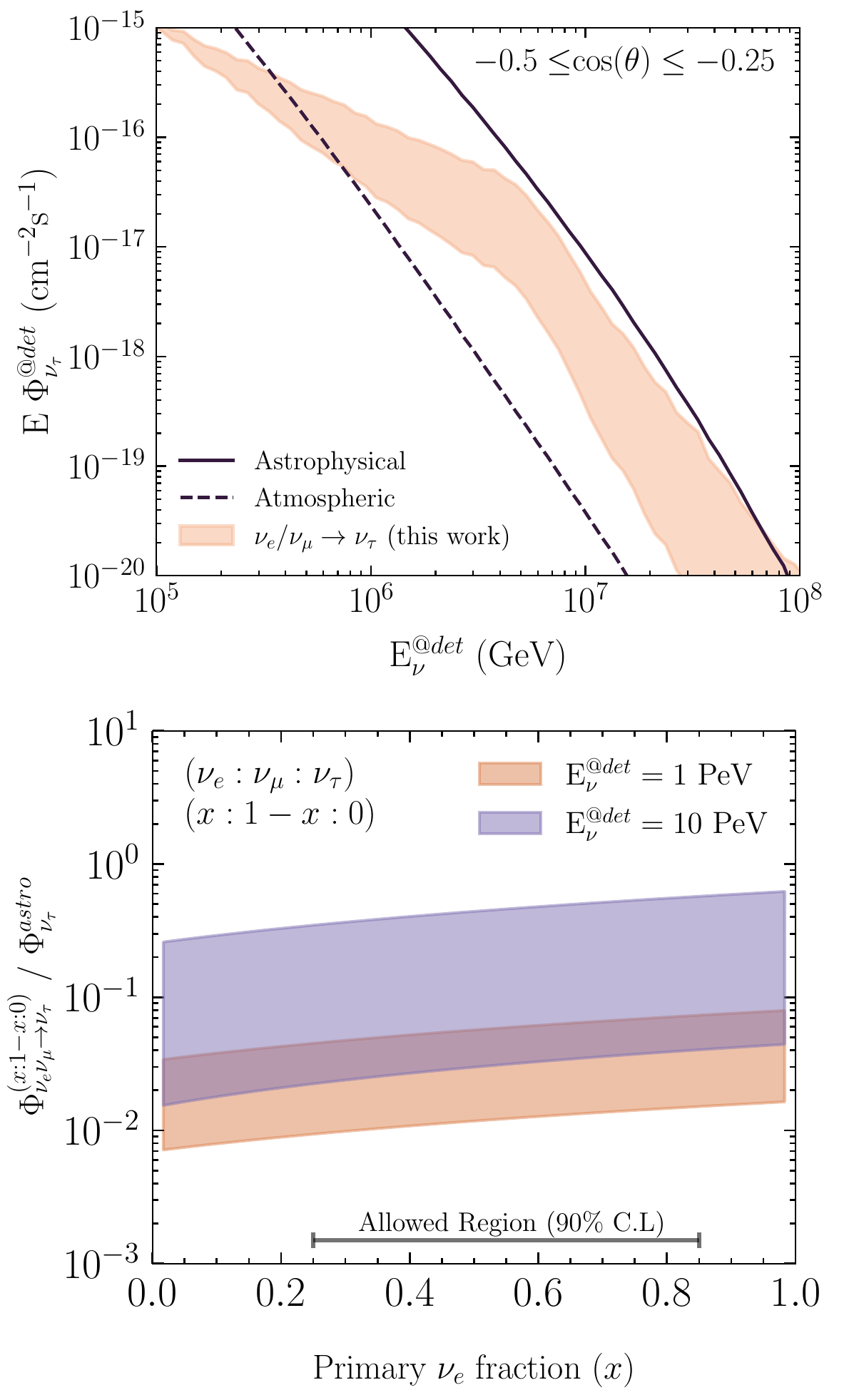}
\caption{
\textbf{\textit{Primary neutrino fluxes and appearing tau-neutrino component.}}
Top: Different components of tau-neutrino flux at the detector integrated over $-0.5<\cos\theta<-0.25$.
Continuous line shows the expected flux using best-fit points from the HESE analysis~\protect\cite{IceCube:2020wum}.
Dashed line shows the atmospheric flux using \texttt{MCeq} (H3a-SIBYLL23C)~\protect\cite{Fedynitch:2015zma}.
The shaded region represents tau neutrinos appearing from the propagation of muon and electron neutrinos, assuming $(2:1:0)$ flavor composition.
Bottom: Ratio of the secondary flux assuming different muon- and electron-neutrino fractions to the astrophysical flux assuming $\left(1:1:1\right)$ at $\SI{1}\PeV$ (salmon) and $\SI{10}\PeV$ (lavender).
The bands represent the uncertainty in the primary spectrum.
The black, horizontal line indicates the 90\% C.L. allowed flavor composition~\protect\cite{IceCube:2018pgc,IceCube:2015gsk}.}
\label{fig:astro}
\end{figure}

In this letter, we estimate the expected flux of tau neutrinos at the detector under two different hypotheses.
First, we assume the best-fit result from HESE under the canonical $\left(1:1:1\right)$ flavor composition.
Second, we use the best-fit results from HESE and northern tracks assuming a zero tau-neutrino component in the primary flux.
\Cref{fig:astro} shows the expected tau-neutrino flux at the detector for events below the horizon ($-0.5<\cos\theta <-0.25$).
Tau neutrinos produced from muon and electron neutrinos alone can contribute as much as $50\%$ of the primary astrophysical flux for energies above $\SI{10}\PeV$.
This fraction mainly depends on the primary spectrum but not on the relative composition of muon and electron neutrinos as shown in~\cref{fig:astro}.
Additionally, the yield of secondary tau neutrinos has an angular dependence since it varies with the column depths that neutrinos travel.
Hence, the relative fraction will be higher for steeper angles.
Consequently, even in a scenario where the astrophysical neutrino flux is only made of muon and electron neutrinos, there is a guaranteed flux of tau neutrinos in the detector, which is larger than the atmospheric component (from the prompt decay of heavy mesons) above $\mathcal{O}(\SI{100}\TeV)$.

\begin{figure}[!h]
\centering
\includegraphics[width=0.4\textwidth]{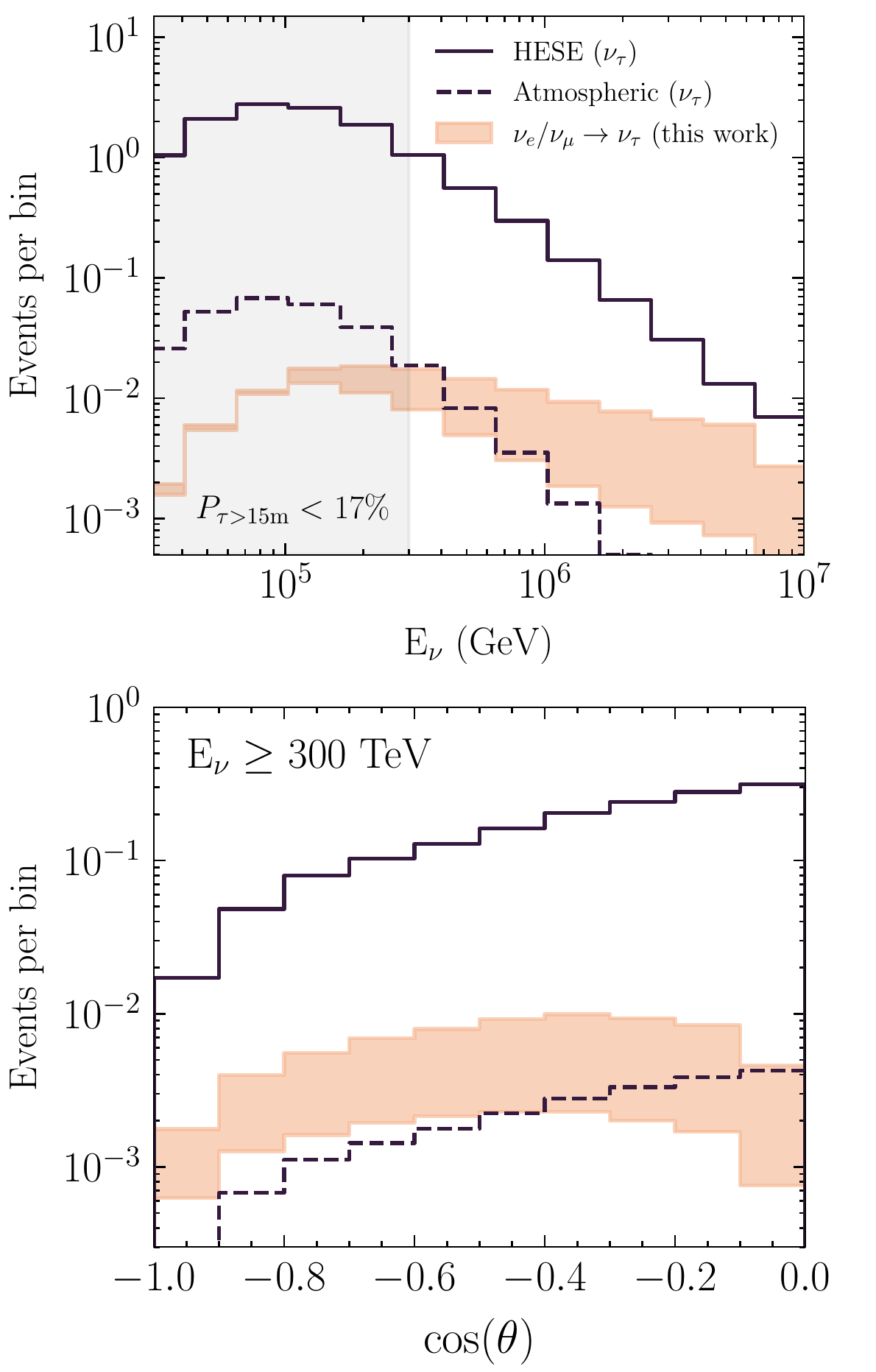}
\caption{
\textbf{\textit{Rates of tau-neutrino events in ten years of HESE.}}
Energy (top) and angular (bottom) distribution of tau-neutrino events that pass the HESE selection.
The angular distribution is shown for events with energies above $\SI{300}\TeV$.
Lines and shaded regions correspond to the fluxes described in~\cref{fig:astro}.
The grey area represents the region where tau-neutrino identification with current reconstruction methods is challenging.
}
\label{fig:rates}
\end{figure}

\textbf{\textit{Impact of tau appearance on discovery of cosmic tau neutrinos ---}}
Here we examine how this additional component manifests in the HESE analysis.
To do so, we combine the expected fluxes at the detector with the publicly available effective areas, assuming the best-fit HESE flux and a democratic primary flavor composition.
\Cref{fig:rates} shows the expected energy and angular distribution of up going tau neutrinos in HESE after ten years of data taking.
Above $\SI{300}\TeV$, we expect 1.6 upgoing tau neutrinos, but only 0.02 events of atmospheric origin.
The expected rate from secondary tau neutrinos is twice that of the prompt atmospheric component.
Although the expected number of events is small, this secondary component is the dominant irreducible background for any astrophysical tau-neutrino search.
\Cref{tab:significane} summarizes the capabilities for rejecting the non-tau cosmic component, assuming that all upgoing tau-neutrino events in the HESE sample above an energy threshold are identified. 
We conclude that this additional tau-neutrino contribution reduces the discovery potential for all energy threshold criteria considered.
A thorough breakdown using realistic tau-neutrino identification criteria as well as a likelihood approach that includes the energy and angular information must be performed by neutrino observatories to fully quantify the impact of this intrinsic background.

\begin{table}[h]
\caption{
\textbf{\textit{Impact of appearing tau-neutrino component on cosmic tau-neutrino discovery.}}
Each row shows a different assumed threshold, as indicated in the first column, to identify tau neutrinos in the HESE analysis.
The second column shows the percentage of tau-neutrino interactions with $E_{\nu}=E_{th}$ producing a tau lepton that travels more than $\SI{15}\meter$~\protect\cite{IceCube:2020abv}.
The third column shows the number of upgoing tau-neutrino events in HESE in ten years.
The significance ($\sigma$) to reject the non-tau cosmic component for different background hypothesis is shown in parenthesis in the fourth and fifth columns when considering atmospheric and appearing-tau backgrounds.
}
\begin{tabular}{l r@{\hskip 0.15in}c@{\hskip 0.15in}c@{\hskip 0.2in}c}
\toprule
$E_{th}$ & $P_{\tau>\SI{15}\m}$ & HESE & Atmos. & $\nu_\mu/\nu_e\rightarrow\nu_\tau$ \\
\colrule
100 TeV &  1\% & 6.63 & 0.13 (6.3) & 0.05-0.11 (6.0-5.7) \\
200 TeV &  9\% & 3.00 & 0.05 (4.3) & 0.03-0.09 (4.1-3.7) \\
300 TeV & 17\% & 1.57 & 0.02 (3.2) & 0.02-0.07 (2.9-2.5) \\
400 TeV & 23\% & 1.12 & 0.01 (2.7) & 0.01-0.06 (2.4-2.1) \\
\botrule
\label{tab:significane}
\end{tabular}
\end{table}

Currently, flavor composition measurements from IceCube are statistically limited due to the size of the detector.
However, the IceCube-Gen2 optical array will be able to detect a sizable amount of PeV neutrinos~\cite{IceCube-Gen2:2020qha}.
The rate of secondary neutrinos due to $W$-boson production, Glashow resonance, and top-quark production is $5-20\%$ above $\SI{10}\PeV$, as shown in~\cref{fig:rates}.
In fact, the fraction of any flavor is affected by these new channels since $W$ bosons can also decay to muon and electron neutrinos.
Therefore, any analysis studying the composition of the astrophysical flux using upgoing events must account for these secondary neutrinos.
Thus, our conclusions are generic and extend beyond the HESE analysis.

\textbf{\textit{Ultra-high-energy neutrinos ---}}
Currently, IceCube measurements extend to $\SI{10}\PeV$ in neutrino energy~\cite{IceCube:2020wum,Abbasi:2021viw,IceCube:2021rpz}. 
At higher energies, many experiments have placed upper limits on the neutrino flux~\cite{IceCube:2018fhm,PierreAuger:2019ens,ANITA:2019wyx,ARA:2015wxq}; however, their effective volumes and Earth's opacity for UHE neutrinos limit their capabilities.
One of the most promising ideas to detect EeV neutrinos is to look for Earth-skimming tau leptons~\cite{Sasaki:2017zwd,GRAND:2018iaj,Wissel:2020sec,Romero-Wolf:2020pzh,POEMMA:2020ykm,Wang:2021zkm}.
Almost horizontal tau neutrinos at energies above $\SI{10}\PeV$ will interact in Earth's crust, producing an energetic tau-lepton, which can emerge in the atmosphere and decay.
Subsequently, electromagnetic radiation from upgoing EASs can be detected at optical and radio wavelengths.
Using this technique, experiments like POEMMA~\cite{Venters:2019xwi}, GRAND~\cite{GRAND:2018iaj}, Trinity~\cite{Wang:2021zkm} and Beacon~\cite{Wissel:2020sec} claim to have significantly better sensitivities than current experiments.
In fact, the projected performance of the aforementioned experiments shows that they would be able to discover the cosmogenic neutrino flux for most UHECRs scenarios.

\begin{figure}[!h]
\centering
\includegraphics[width=0.5\textwidth]{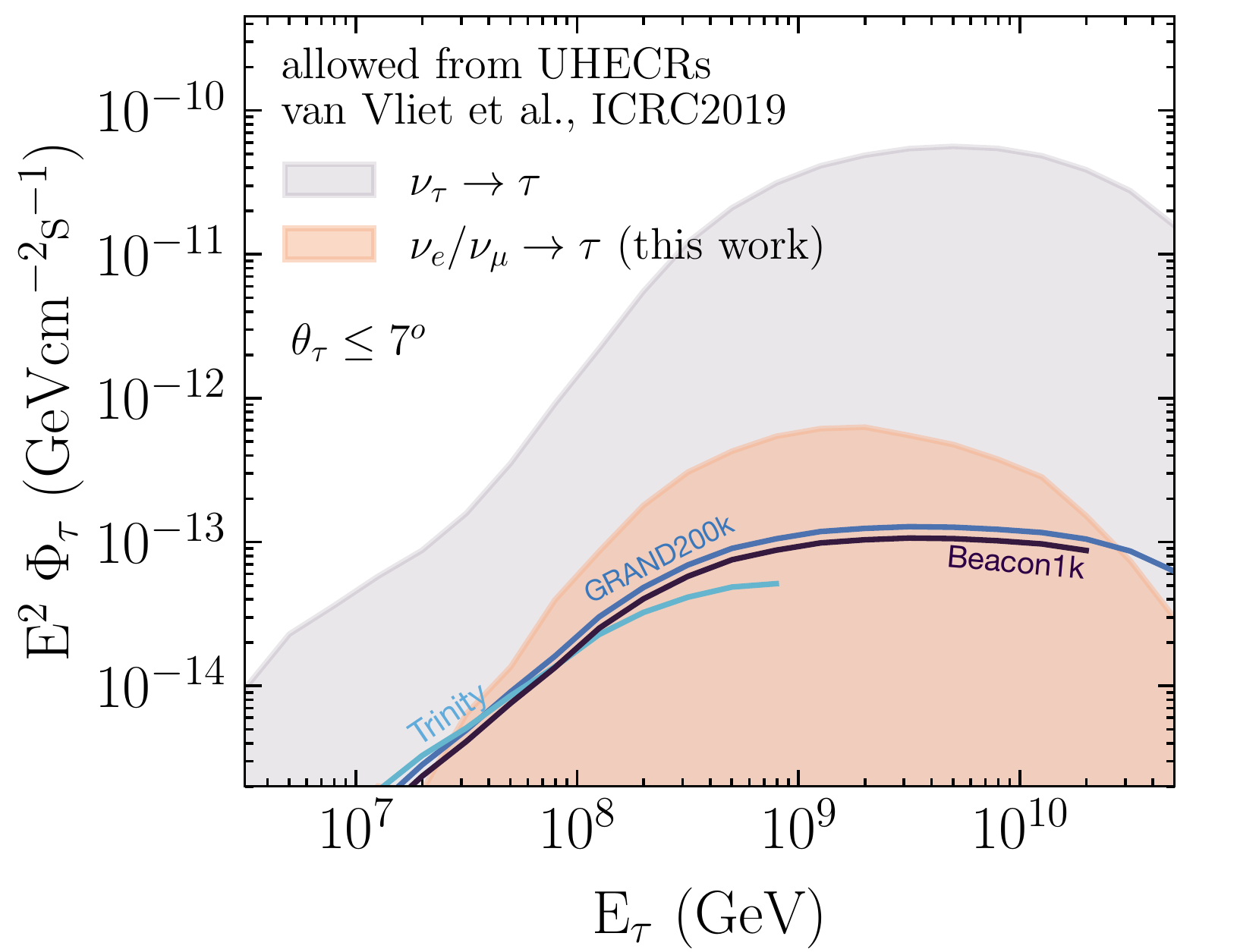}
\caption{
\textbf{\textit{Tau-lepton fluxes arriving at Earth-skimming experiments.}}
Expected flux of emerging tau for Earth-skimming neutrinos ($\theta_{\tau}<7\degree$).
Shaded curves represent the allowed regions derived from a model fit to Auger data~\protect\cite{vanVliet:2019cpl} with (gray) and without (salmon) tau neutrinos in the primary flux.
Solid lines indicate the flux for which Earth-skimming experiments would collect more than 2.44 events in ten years assuming a $E^{-2}$ primary neutrino spectrum.
}
\label{fig:gzk}
\end{figure}

\Cref{fig:gzk} shows the predicted energy spectrum of tau leptons emerging almost horizontally from Earth's surface.
In this calculation, we have assumed a cosmogenic model that includes more than a single UHECR source population~\cite{vanVliet:2019cpl}.
The expected secondary flux coming from muon- and electron-neutrino interactions --- as illustrated in~\cref{fig:scheme} --- is also shown together with the $E^{-2}$ flux for which GRAND, Trinity, and Beacon will observe more than 2.44 events (90\% C.L. upper limit) after ten years of data taking.
Two main conclusions can be derived from the predictions shown in~\cref{fig:gzk}.
First, we expect a non-negligible contribution of emerging taus from a primary flux of muon and electron neutrinos. 
This holds true even if the flavor composition of the cosmogenic flux has no primary tau component.
Thus for optimistic models, GRAND, Trinity, and Beacon will detect taus regardless of primary flavor composition.
Second, when they observe emerging taus, they cannot infer the type of neutrino that produced them.
Therefore, these experiments will not set strong constraints on the normalization of the cosmogenic flux without making assumptions about its flavor composition; \textit{i.e.}, the yield of taus is degenerated between small $\left(1:1:1\right)$ and large $\left(1:1:0\right)$ primary fluxes.

There are some nascent reconstruction techniques to detect EAS induced by muons with these detectors~\cite{Cummings:2020ycz}.
These novel methods would make Earth-skimming experiments sensitive to multiple flavors, allowing them to break the degeneracy previously described.
Similarly, experiments that can detect geo-synchrotron and Askaryan radiation~\cite{Askaryan:1961pfb} --- \textit{e.g.}, IceCube-Gen2~\cite{IceCube-Gen2:2020qha}, RNO-G~\cite{RNO-G:2020rmc} or PUEO~\cite{PUEO:2020bnn} --- would not be affected by this degeneracy because they are equally sensitive to all neutrino flavors.
Therefore, a combined analysis between the mentioned experiments will be fundamental to constrain the normalization and flavor composition of the cosmogenic neutrino flux.

\textbf{\textit{Conclusion and outlook ---}} High-energy muon and electron neutrinos have a non-negligible probability of producing tau neutrinos as they propagate through Earth.
This appearance of tau neutrinos is significant, and yet it has been overlooked in previous literature and experimental analyses.

We have shown that the flux of secondary tau-neutrinos is larger than the prompt atmospheric component above a few hundred TeV.
It is the dominant irreducible background for astrophysical tau-neutrino searches in current and future neutrino telescopes. 
Hence, future flavor composition measurements of the cosmic-neutrino flux must account for it.
The relative impact of this background highly depends on the primary neutrino spectrum, so it is fundamental to constrain its shape and normalization further.

We have proved that Earth-skimming experiments are sensitive to cosmogenic fluxes with multiple source populations of UHECRs, independent of the primary flavor composition.
Even in extreme scenarios without primary tau-neutrinos, there will be a distinguishable flux of emerging tau leptons coming from muon- and electron-neutrino interactions.
Nevertheless, experiments sensitive to a single flavor cannot set constraints on the normalization of the flux since there is a degeneracy between primary and secondary contributions.
We therefore conclude that a symbiotic ecosystem of neutrino telescopes with different flavor identification capabilities is needed to understand the origin of UHE neutrinos.

\textbf{\textit{Acknowledgments ---}}
We thank Jack Pairin for the artistic rendition of our work shown in Fig.~\ref{fig:scheme}.
We also thank Francis Halzen and Dawn Williams for usefull discussions and comments on the manuscript.
We acknowledge engaging discussions on propagation of high-energy neutrinos and their interactions with Sergio Palomares-Ruiz, Hallsie Reno, Rhorry Gauld, Juan Rojo and Aart Heijboer. 
AG acknowledges support from the European Union’s H2020-MSCA Grant Agreement No.101025085.
CAA, IS, and PZ are supported by the Faculty of Arts and Sciences of Harvard University.
Additionally, CAA thanks the Alfred P. Sloan Foundation for their support.

\bibliographystyle{apsrev4-1}
\bibliography{tau-no-tau}

\clearpage
\pagebreak
\appendix
\onecolumngrid
\ifx \standalonesupplemental\undefined
\setcounter{page}{1}
\setcounter{figure}{0}
\setcounter{table}{0}
\setcounter{equation}{0}
\fi
\renewcommand{\thepage}{Supplemental Methods and Tables --- S\arabic{page}}
\renewcommand{\figurename}{SUPPL. FIG.}
\renewcommand{\tablename}{SUPPL. TABLE}
\renewcommand{\theequation}{A\arabic{equation}}
\clearpage
\begin{center}
\textbf{\large Supplemental Material}
\end{center}

\textbf{\textit{Neutrino Cross Sections}}: The structure functions for neutrino-nucleon deep inelastic scattering have been implemented in GENIE.
In this study, the HEDIS-CSMS model (\texttt{GHE19\_00b}) is used.
This calculation uses the NLO HERA1.5 PDF set~\cite{Cooper-Sarkar:2010yul} and uses expressions for the massless coefficient functions at NLO.
This is the preferred model in IceCube's analyses.
Additionally, the $W$-boson production and Glashow resonance are computed using the differential cross sections as described in~\cite{Garcia:2020jwr}.

The following figure shows the total cross section for channels in which a tau neutrino is produced from $\SI{10}\TeV$ to $\SI{100}\EeV$.
For comparison, the deep inelastic scattering contribution is also shown.

\begin{figure}[!h]
\centering
\includegraphics[width=0.5\textwidth]{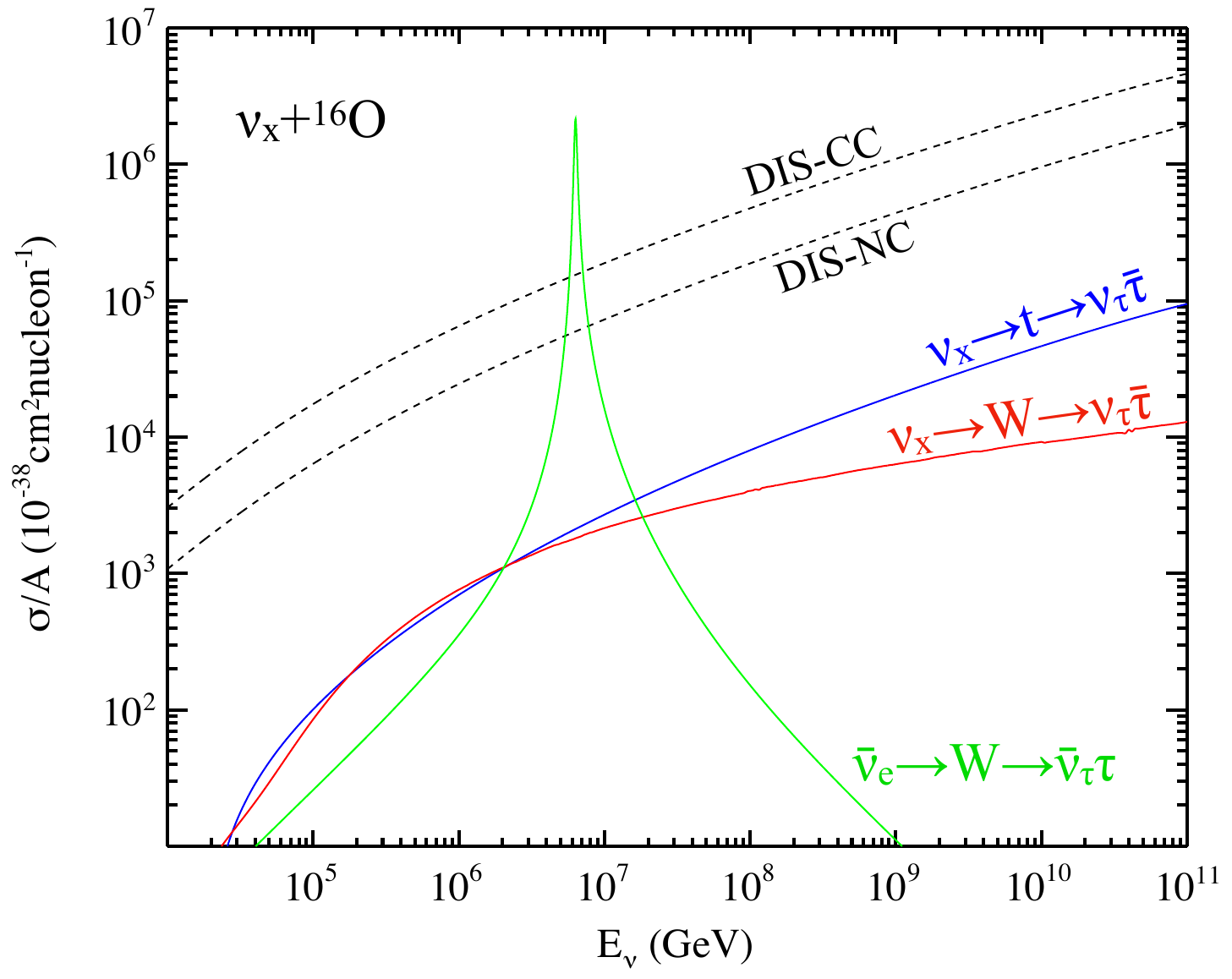}
\caption{
Neutrino-oxygen cross sections as a function of the neutrino energy.
Continuous lines represent the interaction channels in which a $\nu_\tau+\tau$ pair is produced: Glashow resonance (blue), $W$-boson production (red), and top-quark production (green).
The deep inelastic scattering contribution is shown with a dashed line (summing CC and NC).
}
\label{fig:xsec}
\end{figure}

\clearpage

\textbf{\textit{Flux of Secondary Tau Neutrinos in different angular regions:}} The following figure shows the expected flux of tau neutrinos at the detector for different components in four angular regions. 
In addition, the relative fractions of secondary tau neutrinos to the astrophysical component at 1~PeV and 100~PeV are shown.

\begin{figure}[!h]
\centering
\includegraphics[width=1\textwidth]{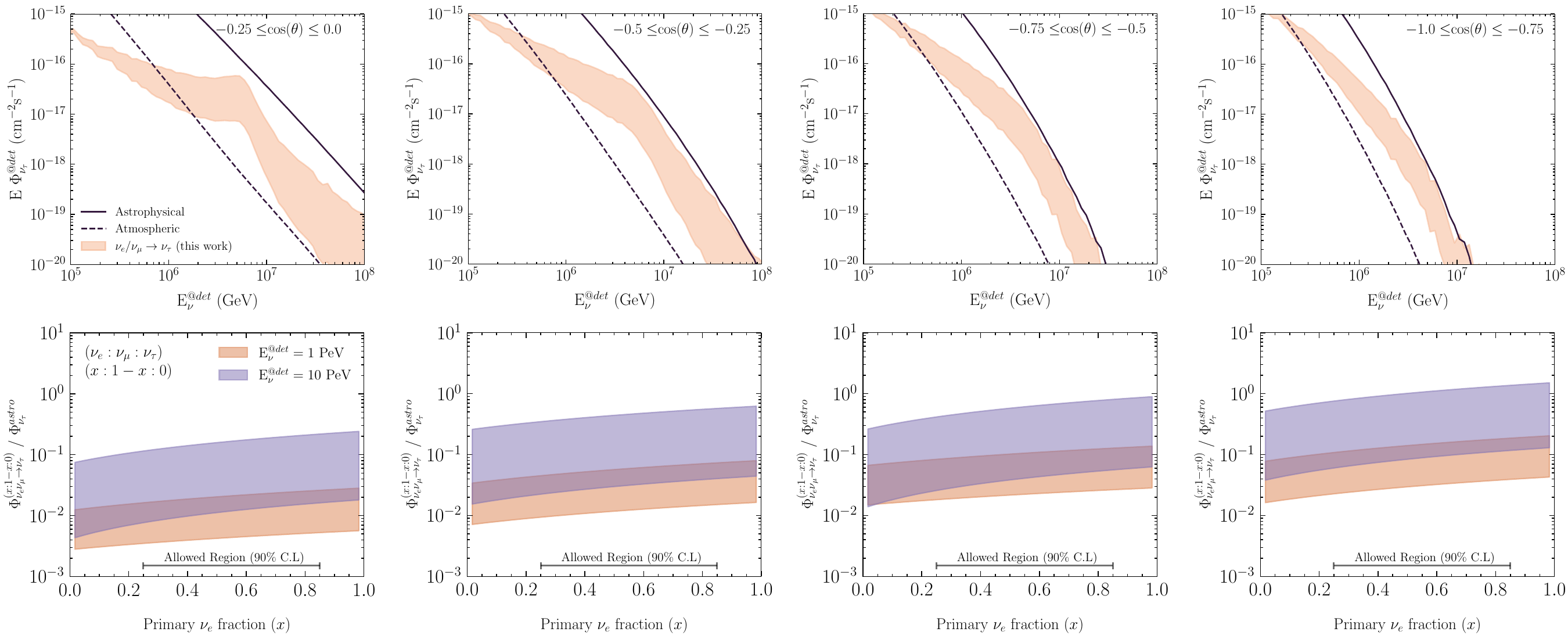}
\caption{Top: Components of tau-neutrino flux at the detector integrated over different angular regions.
Continuous line shows the expected flux using best-fit points from the HESE analysis~\protect\cite{IceCube:2020wum}.
Dashed line shows the atmospheric flux using \texttt{MCeq} (H3a-SIBYLL23C)~\protect\cite{Fedynitch:2015zma}.
The shaded regions show the secondary tau neutrinos contribution produced by the propagation of muon and electron neutrinos assuming the best fits from IceCube's analysis~\protect\cite{IceCube:2020wum,Stettner:2019tok} and $\left(2:1:0\right)$ flavor composition.
Bottom: Ratio of  the secondary flux assuming different muon- and electron-neutrino fractions to the HESE's best-fit flux assuming $\left(1:1:1\right)$ at $\SI{1}\PeV$ (red) and $\SI{10}\PeV$ (blue).
The bands represent the uncertainty in the primary spectrum.
The black, horizontal line indicate the combined 90\% C.L. from different flavor composition measurements~\cite{IceCube:2018pgc,IceCube:2015gsk}.}
\label{fig:suppl2}
\end{figure}

\end{document}